\documentclass[12pt]{iopart}

\pdfoutput=1
\usepackage{graphicx}
\usepackage{floatrow} 
\begin{document}

\title[Electronic band structures in 2D materials]{Electronic structures near unmovable nodal points and lines in two-dimensional materials}

\author{V Damljanovi\'c$^1$ and N Lazi\'c$^2$}

\address{$^1$Institute of Physics Belgrade, University of Belgrade, Pregrevica 118, 11000 Belgrade, Serbia}
\ead{damlja@ipb.ac.rs}
\address{$^2$NanoLab, Faculty of Physics, University of Belgrade, Studentski Trg 12, 11001 Belgrade, Serbia}
\vspace{10pt}
\begin{indented}
\item[]December 2022
\end{indented}

\begin{abstract}
Unmovable nodal points (UNP) and lines (UNL) are band crossings which positions in the Brillouin zone are unaltered by symmetry preserving perturbations. Not only positions but also the band structure in the vicinity is determined by the little group of a wave vector and it's irreducible (co)representations. In this paper we give the full set of electronic dispersions near all UNPs and UNLs in non-magnetic, quasi two-dimensional (2D) materials both with and without spin-orbit coupling (SOC). Analysis of all layer gray single and double groups gives nineteen different quasiparticles, great majority of which are unavoidable for a 2D material which belongs to certain layer groups. These include Weyl and Dirac nodal lines, dispersions with quadratic or cubic splitting, anisotropic Weyl and Dirac cones which orientation can be varied by \emph{e.g.} strain \emph{etc}. We indicated qusiparticles that are robust to SOC. For convenience, our results are concisely presented graphically - as a map, not in a tabular, encyclopedia form. They may be of use as checkpoints or for fitting of experimentally (via \emph{e.g.} ARPES) and numerically obtained electronic band structures as well as for deeper theoretical investigations.
\end{abstract}

%
% Uncomment for keywords
%\vspace{2pc}
%\noindent{\it Keywords}: XXXXXX, YYYYYYYY, ZZZZZZZZZ
%
% Uncomment for Submitted to journal title message
%\submitto{\JPA}
%
% Uncomment if a separate title page is required
%\maketitle
% 
% For two-column output uncomment the next line and choose [10pt] rather than [12pt] in the \documentclass declaration
%\ioptwocol
%

\section{Introduction}

The Fermi-Dirac distribution dictates that most of the material properties depend on the electronic dispersions shape near the Fermi energy in the Brillouin zone (BZ). On the other hand, these shapes are solely determined by the crystal symmetry and, for non-magnetic materials, the time-reversal symmetry (TRS). Presence of Dirac cones around the BZ $K-$points of the honeycomb lattice is the most notable example. Experimental discovery of the electric field effect in graphene, fabrication of graphene's three dimensional (3D) analogs accompanied with the explanation of certain band structure properties using topological laws, brought topological materials into the focus of solid-state science \cite{TopMat1, TopMat2, TopMat3}. Non-trivial band topology, such as non-zero Berry phase, originates from singularities of eigenvectors in the reciprocal space. Such singularities often appear at points where two or more bands touch.

In two-dimensional (2D) materials, bands can touch at isolated points or at lines. Band touchings caused by the increased dimensionality of corepresentation (correp) of little group do not move if the Hamiltonian parameters change, as long the symmetry is preserved. Such touchings we call unmovable nodal points (UNP) and lines (UNL). Weyl/Dirac nodal lines (WNL)/(DNL) are examples of UNL, which presence in a band structure leads to interesting physical properties \cite{VeLiR1, VeLiR2, VeLiR3, VeLiR4}. These include graphene-like density of states (DOS) \cite{ToNoSe1, ToNoSe2} and induction of drumhead states which flat bands may cause strong electron correlations \cite{VeLiR2, Volo}. In 3D, WNL and DNL were determined in achiral gray double space groups \cite{3DSym2}, in 230 ordinary/gray single/double space groups \cite{3DSym1}, in certain black-and-white magnetic space groups \cite{3DSym3} and in the presence and absence of spin-orbit coupling (SOC) \cite{3DSym4}. Refs. \cite{2DDNL1, 2DDNL2} report DNL in the presence of SOC in 2D. On the other hand, full enumeration of layer groups' UNPs hosting fully linear dispersion in non-magnetic 2D materials is completed task \cite{Mi22}. In Refs. \cite{Mi22, Ja17, Mi20} the Hamiltonians are distinguished from each other by the form of their eigenvalues. The term Dirac was reserved for four-component Hamiltonians with graphene-like, isotropic or anisotropic eigenvalues \cite{Mi22, Ja16, Ja16Ad}. There are several physical reasons for this distinction, \emph{e.g.} the non-vanishing DOS of some dispersion \cite{Ja17, Mi20} near zero energy. On the other hand, analysis of all dispersions other than linear in 2D materials has not been reported thus far \cite{Rev2D}. Recently published reports for (3D) space groups \cite{Enci1, Enci2, Enci3, Enci4} could be (in principle) used for layer groups, if one subduces the results for corresponding space groups to 2D BZ. Ref. \cite{Enci2D} suggest that this method is indeed applied for layer groups but, in all fairness, it is published (at the moment of submission of this manuscript) without the Supplementary Material which contains the results announced in the main part of the paper. Ref. \cite{Enci5} has classified all massless excitations in 2D and 3D.

In this paper we calculated exact shapes of electronic bands near all high-symmetry points (HSP) and lines (HSL) in the BZ of all layer groups, including cases without and with SOC, in the presence of TRS. It turns out that in the Taylor expansion of Hamiltonians, the relevant terms are up to maximally third order in momentum. We found in total nineteen quasiparticles, almost none of which were investigated in the present literature. Our results are presented graphically, as an atlas (map) rather than in a tabular, encyclopedia form, and fit into two figures and a few pages of text/formulas. For all symmetry constrained Hamiltonians, the eigenvalues are given. We single out quasiparticles that are not destroyed by inclusion of SOC. 

\section{Method}

To find all possible HSPs and HSLs with multidimensional correps, one has to search through all little groups of all gray layer groups - 80 for spinless and 80 for spinful cases. Unlike correps of space groups, which were tabulated and made available long time ago \cite{BCSSG, BCSDSG}, layer group representations became publicly available only recently \cite{DGSITE}. Ref. \cite{DGSITE} was a starting point of our search.

Symmetry adapted, low-energy Hamiltonians were derived from the following formula \cite{Mi22, Manjes12}:

\begin{equation}
\label{crsym}
\hat{H}\left(\mathbf{k}_0+\mathbf{q}\right)=\hat{D}^{\dag}\left(g\right)\hat{H}\left(\mathbf{k}_0+\hat{h}'\mathbf{q}\right)\hat{D}\left(g\right),
\end{equation}
where $\mathbf{k}_0$ denotes HSP or belongs to HSL, $g=\left\{\hat{h}\right|\left.\mathbf{\tau}_{\hat{h}}\right\}$ is an element (in Seitz notation) of the little group $G_{\mathbf{k}_0}$ ($\hat{h}'\mathbf{k}_0$ is equivalent to $\mathbf{k}_0$), $\hat{h}'$ is an operator reduction of $\hat{h}$ to 2D BZ and $\hat{D}\left(g\right)$ is the irrep matrix that corresponds to $g$. If $-\mathbf{k}_0$ belongs to the star of $\mathbf{k}_0$, the full symmetry group of $\mathbf{k}_0$ is $G_{\mathbf{k}_0}+\theta g_0 G_{\mathbf{k}_0}$, where $\theta$ denotes time-reversal and the element $g_0=\left\{\hat{h}_0\right|\left.\mathbf{\tau}_{\hat{h}_0}\right\}$ does not belong to $G_{\mathbf{k}_0}$, if $\mathbf{k}_0$ is not the time reversal invariant momentum ($\hat{h}'_0\mathbf{k}_0$ is equivalent to $-\mathbf{k}_0$). The additional constrain on contribution $\hat{H}'$ appearing in the Taylor expansion $\hat{H}\left(\mathbf{k}_0+\mathbf{q}\right)\approx\hat{H}\left(\mathbf{k}_0\right)+\hat{H}'\left(\mathbf{q}\right)$ is \cite{Mi22, Manjes12}:
\begin{equation}
\label{trs}
\hat{H}'^*\left(\mathbf{q}\right)=\hat{R}^{\dag}\left(\theta g_0\right)\hat{H}'\left(-\hat{h}_0'\mathbf{q}\right)\hat{R}\left(\theta g_0\right),
\end{equation}
where $\hat{R}$ is correp matrix (linear part - without complex conjugation) that corresponds to $\theta g_0$. As opposed to dispersions, which are invariants uniquely determined by the symmetry, the Hamiltonians can be changed by different choice of basis wave-functions. However, once the particular Hamiltonian $\hat{H}(\mathbf{q})$ in the vicinity of $\mathbf{k}_0$ is chosen, the Hamiltonian $\hat{H}_p(\mathbf{q})$ near $\mathbf{k}_p\equiv\hat{h}\mathbf{k}_0$ from the star of $\mathbf{k}_0$, is fixed by the following formula:
\begin{equation}
\label{star}
\hat{H}_p\left(\mathbf{q}\right)=\hat{R}\left(g_p^{-1}g\right)\hat{H}\left(\hat{h}^{-1}\mathbf{q}\right)\hat{R}^{\dag}\left(g_p^{-1}g\right),
\end{equation}
where $g_p=\left\{\hat{h}_p|\mathbf{\tau}_p\right\}$ is the representative of the coset that contains the element $g=\left\{\hat{h}|\mathbf{\tau}\right\}$ of the layer group, and $\hat{R}$ is the correp matrix of the little group of $\mathbf{k}_0$. In (\ref{star})
one has to use the very same correp that gives $\hat{H}(\mathbf{q})$ near $\mathbf{k}_0$, not just an equivalent one. Similarly, for dispersions around $\mathbf{k}_p$:
\begin{equation}
\label{disT}
E_p(\hat{h}\mathbf{q})=E(\mathbf{q}).
\end{equation}
Also, dispersions around two points in the reciprocal space, that differ by a vector from the reciprocal lattice are identical. All necessary matrices are available in PolSym code in addition to the web page \cite{DGSITE, Pol15}. 

To avoid unnecessary complication, we do not list here all Hamiltonians that correspond to each particular correp from \cite{DGSITE}, since many of them can be transformed into each other by the similarity transformation. If two $n$-component Hamiltonians $\hat{H}_1$ and $\hat{H}_2$, have all eigenvalues the same, then $\hat{H}_2=\hat{U}\hat{H}_1\hat{U}^{\dag}$, with $\hat{U}=\sum_{l=1}^n\left|\Psi_l\right\rangle\left\langle \Phi_l\right|$ being an unitary matrix and $\left|\Psi_l\right\rangle$ ($\left|\Phi_l\right\rangle$) the normalized eigenvector of $\hat{H}_2$ ( $\hat{H}_1$) that corresponds to the eigenvalue $E_l$, $(l=1,2,...,n)$. Therefore, of all Hamiltonians having the same eigenvalues, only one representative is shown.

For distinction between anisotropic Dirac and Weyl dispersions which orientation depends on the parameters of the model, and the ones with fixed orientation, we use the following theorem. The quadratic form $f=c_1q_1^2+c_3q_1q_2+c_2q_2^2=\left\langle \mathbf{q}\right|\hat{A}\left|\mathbf{q}\right\rangle$, can be reduced to $f=\lambda_1q_1'+\lambda_2q_2'$, where $q_j'=\left\langle \lambda_j\right.\left|\mathbf{q}\right\rangle$, while $\lambda_j$ and $\left|\lambda_j\right\rangle$ are eigenvalues and eigenvectors of the matrix $\hat{A}=\left(
\begin{array}[c]{cc}
	c_1 & c_3/2 \\
	c_3/2 & c_2
\end{array}
\right)$,
$(j=1, 2)$. The $\lambda_j$ and $\left|\lambda_j\right\rangle$ are well known and won't be presented here. The cross-section of $f$ is an ellipse whose orientation (in principle) depends on parameters $c_1$, $c_2$ and $c_3$. The orientation of the ellipse is fixed if eigenvectors are independent on $c_1$, $c_2$ and $c_3$. This is the case if either $c_3=0$ or:
\begin{equation}
\label{uslov}
c_2-c_1=uc_3,
\end{equation}
where $u$ is a real constant, independent of $c_1$, $c_2$ and $c_3$.

\section{Results}
\label{Res}
\begin{figure}
\includegraphics[width=\textwidth]{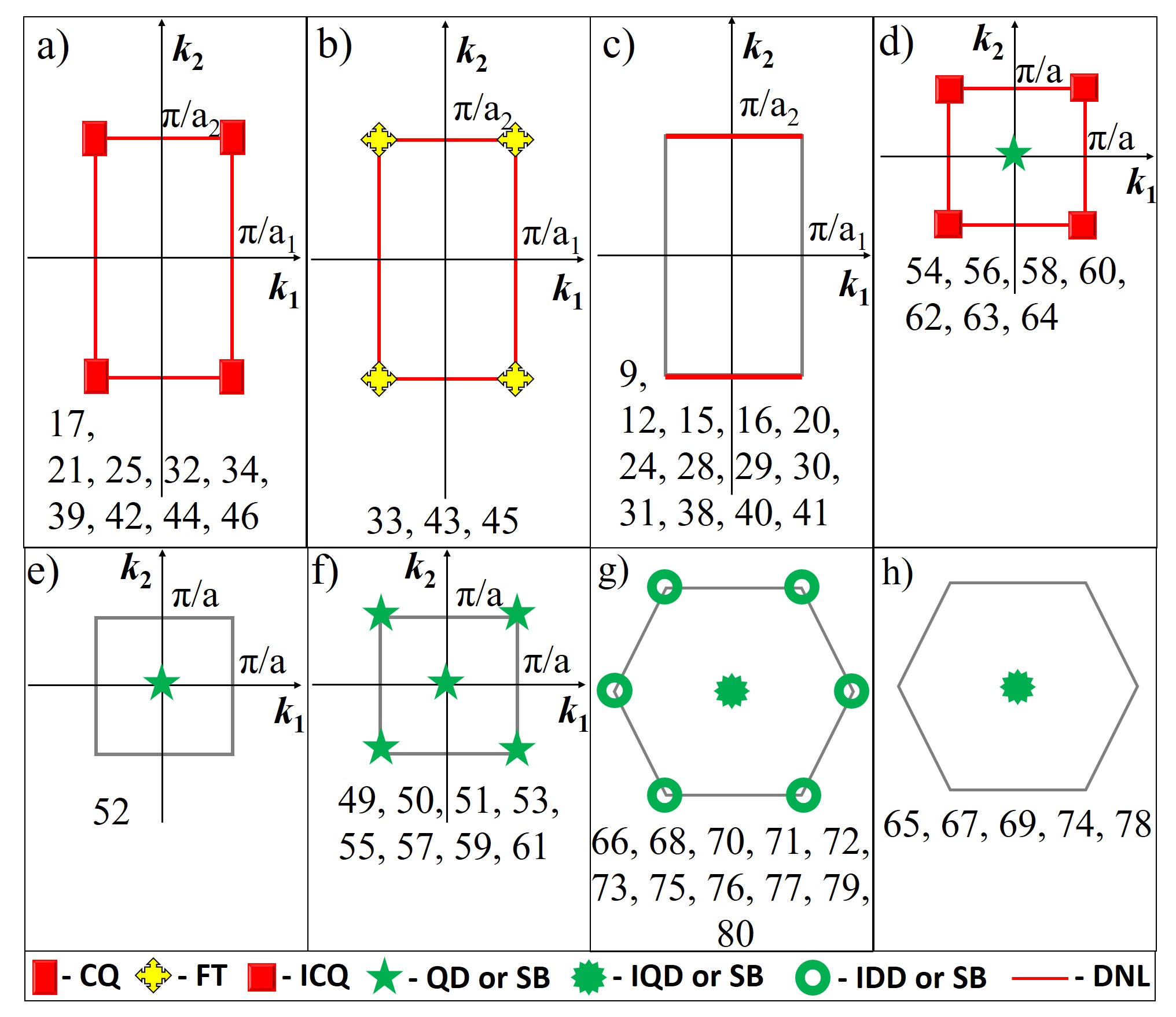}%
\caption{\label{Figu1} Dispersions near UNP and UNL in the BZ of layer groups without SOC (notation as in \cite{vole}). Dispersion acronyms in the lowest panel and the corresponding subsections in the main text are: DNL - Dirac nodal line - \ref{DlnDHs}, FT - fortune teller - \ref{FTs}, IDD - isotropic Dirac - \ref{IsoDDs}, CQ - corner quadratic - \ref{kvads}, ICQ - isotropic corner quadratic - \ref{iskvads}, QD - quadratic - \ref{cukvs}, IQD - isotropic quadratic - \ref{kvs} and SB - simple band - \ref{jdkvs}. The green color denotes more than one possibility for a dispersion around a given BZ point. Other colors denote spinful degeneracy at $\mathbf{q}=0$: red (4-fold) and yellow (8-fold).}
\end{figure}

\begin{figure}
\fl \kern-2em \kern-2em \floatbox[{\capbeside\thisfloatsetup{capbesideposition={left,top},capbesidewidth=0.4\textwidth}}]{figure}[\FBwidth]
{\caption{Dispersions near UNP and UNL in the BZ of layer groups with SOC (superscript $D$ is omitted, notation as in \cite{vole}). Dispersion acronyms in the lowest panel and the corresponding subsections in the main text are: WNL - Weyl nodal line - \ref{WlnDHs}, DNL - Dirac nodal line - \ref{DlnDHs}, IWD - isotropic Weyl - \ref{Dirs}, AWDv -anisotropic Weyl with variable orientation - \ref{avDirs}, AWDf -anisotropic Weyl with fixed orientation - \ref{afDirDs}, FT - fortune teller - \ref{FTs}, PF-poppy flower - \ref{PF1s}, ADDv -anisotropic Dirac with variable orientation - \ref{DDvs}, ADDf -anisotropic Dirac with fixed orientation - \ref{DDfs}, IDD - isotropic Dirac - \ref{IsoDDs}, CQ - corner quadratic - \ref{kvads}, ICQ - isotropic corner quadratic - \ref{iskvads}, SB - simple band - \ref{jdkvs}, CD - cubic - \ref{cps} and TC - triangular cubic - \ref{hexas}. The green color denotes more than one possibility for a dispersion around a given BZ point. Other colors denote spinful degeneracy at $\mathbf{q}=0$: blue (2-fold) and red (4-fold).}\label{Figu2}}
{\includegraphics[width=0.7\textwidth]{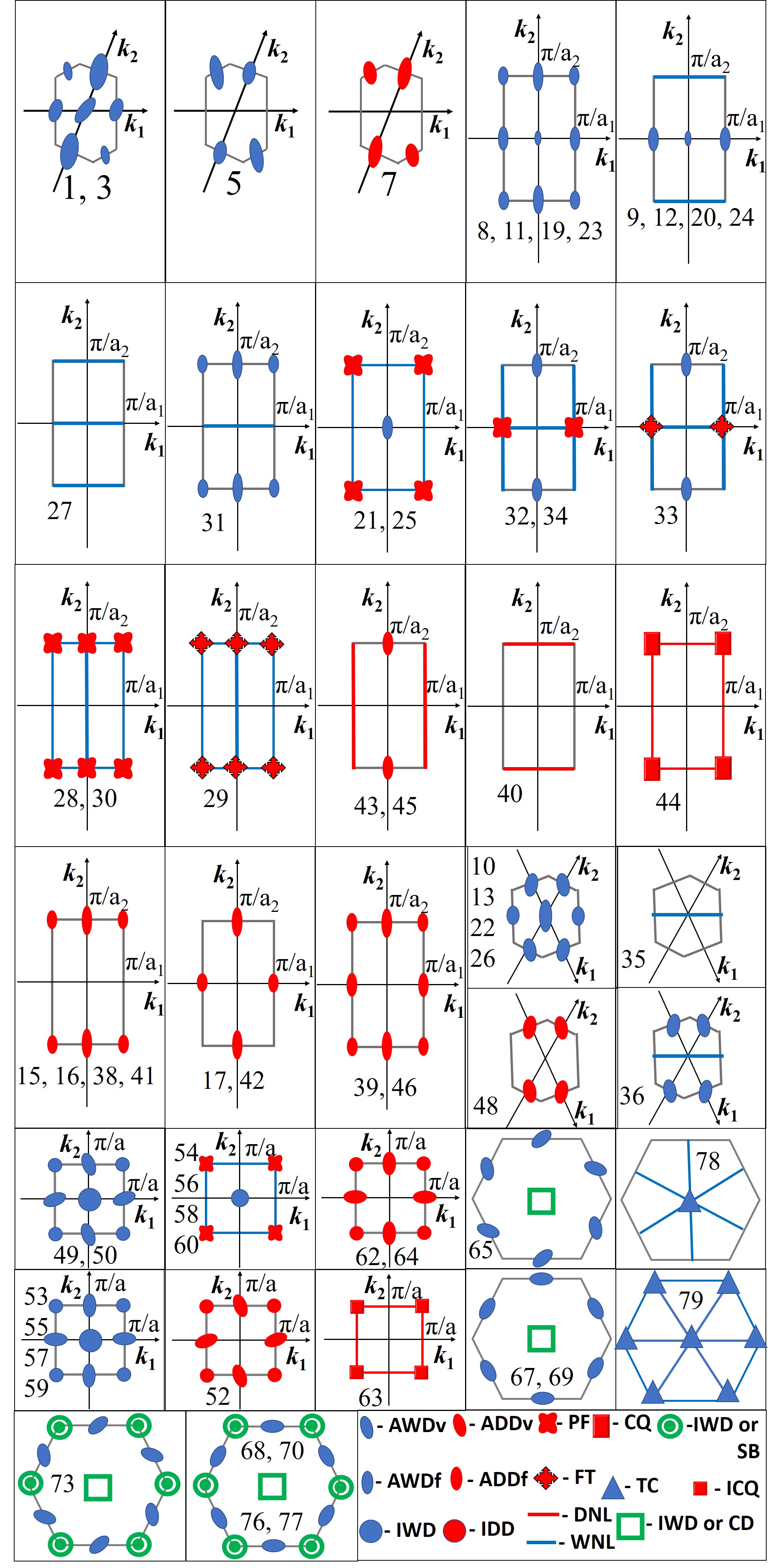}}
\end{figure}

All groups and all BZ points and lines hosting electronic dispersions in the absence of SOC are shown in the Figure \ref{Figu1}, those with SOC are shown in the Figure \ref{Figu2}. Dispersions are denoted with different symbols, as indicated in the captions and explained in the subsections below. Dispersions of the same type but with different parameters are denoted with the same symbols but of different size. 
We used Eq. (\ref{disT}) for graphical presentation of dispersions in the whole BZ. Primitive vectors of reciprocal lattice are $\mathbf{k}_1$ and $\mathbf{k}_2$. For oblique and rectangular-p groups with all non-symmorphic elements having fractional translations parallel to one direction (for group 45), $\mathbf{k}_1$ is perpendicular to that direction (to the symmorphic reflection plane). For rectangular-c groups $35^D$ and $36^D$, WNL is along symmorphic axis of order two. For groups 27 and 34 (32, 33 and 43), $\mathbf{k}_1$ is along axis (screw axis) of order two. This convention is of use when directions along $\mathbf{k}_1$ and $\mathbf{k}_2$ differ significantly in dispersions they host. For rectangular-c groups in Figure \ref{Figu2}, BZ is shown for the angle between primitive vectors greater than $\pi/2$. The results are equally valid if that angle is sharp, just the BZ border would be rotated. Groups not shown in Figure \ref{Figu1} (Figure \ref{Figu2}) do not have UNP and/or UNL in the case without (with) SOC.

The studies are performed for linear, quadratic and, when it was relevant, for the third order in $\mathbf{q}$, and all together we single out 19 types of dispersions. In the following, they are described in more detail. Letters $a$ and $b$ denote real numbers, $z$ denotes complex numbers, the superscript $D$ denotes layer double group, $\hat{\sigma}_1$, $\hat{\sigma}_2$, $\hat{\sigma}_3$ ($\hat{\sigma}_0$) are Pauli matrices (is the two-dimensional unit matrix), while $q_1$ and $q_2$ are projections of $\mathbf{q}$ along orthonormal vectors $\mathbf{e}_1$ and $\mathbf{e}_2$ respectively. 

\subsection{Nodal line dispersions}

\subsubsection{Weyl nodal line (WNL)}
\label{WlnDHs}

WNL is line of double spinful degeneracy which splits linearly in the direction perpendicular to the line. It appears only in SOC case and it is denoted with blue lines in the Figure \ref{Figu2}. The dispersion and the Hamiltonian are:
\numparts
\label{Wln}
\begin{equation}
\label{WlnD}
E_{1,2}=E_0+a_1q_{||}\pm c\left|q_{\bot}\right|,
\end{equation}
\begin{equation}
\label{WlnH}
\hat{H}=\left(E_0+a_1q_{||}\right)\hat{\sigma}_0+q_{\bot}\sum_{j=1}^3b_j\hat{\sigma}_j,
\end{equation}
\endnumparts
where $c=\sqrt{\sum_{j=1}^3b_j^2}>0$ and $q_{\bot}$ ($q_{||}$) is projection of $\mathbf{q}$ perpendicular to (along) the line. For groups $20^D$, $21^D$, $24^D$, $25^D$, $54^D$, $56^D$, $58^D$ and $60^D$, $b_3=0$. For groups $27^D$, $28^D$, $29^D$(WNLs perpendicular to $\mathbf{k}_1$), $30^D$, $31^D$, $32^D$, $33^D$(WNL perpendicular to $\mathbf{k}_2$), $34^D$, $35^D$, $36^D$, $78^D$ and $79^D$, $b_3=b_2=0$. For points in the middle and at both ends of WNL, if not occupied by any other dispersion, the following substitution should be made in the last two equations above: $E_0+a_1q_{||}\rightarrow E_0+a_1q_{||}^2$. In these points, one of the bands achieves extremum at $\mathbf{q}=0$ with the effective mass: $\hat{m}_{eff}=\left(
\begin{array}[c]{cc}
	0 & 0 \\
	0 & m_{||}
\end{array}
\right)$, which means that excitations are massless along the direction perpendicular to the WNL.

\subsubsection{Dirac nodal line (DNL)}
\label{DlnDHs}
DNL is a line of 4-fold spinful degeneracy, which splits into two double degenerate bands in the direction perpendicular to the line. DNL appears both in SOC and non-SOC case and it is denoted by a red line. Dispersion is:
\numparts
\label{DlnDH}
\begin{equation}
\label{DlnD}
E_{1,2,3,4}=E_0+aq_{||}\pm c\left|q_{\bot}\right|,
\end{equation}
where $q_{\bot}$ ($q_{||}$) is projection of $\mathbf{q}$ perpendicular to (along) the line. For layer single groups, the total Hamiltonian is $\hat{H}_{tot}=\hat{\sigma}_0\otimes\hat{H}$, with $\hat{H}$ given by (\ref{WlnH}), while $E_1^{\uparrow}=E_1^{\downarrow}=E_1=E_2$, $E_2^{\uparrow}=E_2^{\downarrow}=E_3=E_4$ and $c=\sqrt{\sum_{j=1}^3b_j^2}$. For groups 15, 16, 17, 20, 21, 24, 25, 40, 43(DNLs perpendicular to $\mathbf{k}_1$), 44, 45(DNLs perpendicular to $\mathbf{k}_1$), 54, 56, 58, 60 and 63, $b_3=0$. For groups 28, 30, 31, 32, 33(DNLs perpendicular to $\mathbf{k}_2$), 34, 38, 39, 41, 42, 43(DNLs perpendicular to $\mathbf{k}_2$), 45(DNLs perpendicular to $\mathbf{k}_2$), 46, 62 and 64, $b_3=b_2=0$. For points in the middle and at both ends of DNL, if not occupied by any other dispersion, the following substitutions should be made in the equations for Hamiltonian (\ref{WlnH}) and dispersion (\ref{DlnD}): $E_0+a_1q_{||}\rightarrow E_0+a_1q_{||}^2$ and $b_2=b_3=0$. For layer double groups, the Hamiltonian is: 
\begin{equation}
\label{DlnH}
\hat{H}=(E_0+aq_{||})\hat{I}_4+q_{\bot}\left(
\begin{array}[c]{cccc}
	b & 0 & z & 0 \\
	0 & -b & 0 & -z \\
	z^* & 0 &-b & 0 \\
	0 & -z^* & 0 & b
\end{array}
\right),
\end{equation}
\endnumparts
with $c=\sqrt{b^2+|z|^2}$. For points in the middle and at both ends of DNL, if not occupied by any other dispersion, the following substitutions should be made in (\ref{DlnH}): $E_0+a_1q_{||}\rightarrow E_0+a_1q_{||}^2$ (endpoints of $43^D$ and $45^D$) and (in addition) $b=0$ (for remaining such points).

\subsection{Linear dispersions}

\subsubsection{Isotropic Weyl dispersion (IWD)}
\label{Dirs}
IWD consists of two spinful non-degenerate isotropic cones touching each other at $\mathbf{q}=0$ and it appears only in the case with SOC. IWD is denoted with a blue circle in Figure \ref{Figu2} (when no other dispersion is possible around that point), with a green ring with a dot inside (when simple band (SB) described latter is an alternative), or with a green square (when cubic dispersion (CD) described latter is an alternative). The dispersion and Hamiltonian are:
\numparts
\label{Dir}
\begin{equation}
\label{DirD}
E_{1,2}=E_0\pm |z|\left|\mathbf{q}\right|,
\end{equation}
\begin{equation}
\label{DirH}
\hat{H}=E_0\hat{\sigma}_0+\left(
\begin{array}[c]{cc}
	0 & z(q_1-iq_2)  \\
	z^*(q_1+iq_2) & 0
\end{array}
\right),
\end{equation}
\endnumparts
for groups $54^D$, $56^D$, $58^D$, $60^D$, $z=(1-i)b_1$, for groups $67^D$, $69^D$, $z=(1+i\sqrt{3})b_1$, for the remaining groups $z=b_1$, except for $49^D$, $50^D$, $65^D$, $73^D$, where $z$ remains a complex number.

\subsubsection{Anisotropic Weyl dispersion with variable orientation (AWDv)}
\label{avDirs}
AWDv consists of two anisotropic, spinful non-degenerate cones and appears only when SOC is included. The cross section of the cones is an ellipse whose orientation (major axis direction) changes when the parameters of the dispersion change. It is represented by a blue ellipse in Figure \ref{Figu2}, such that its orientation is neither strictly horizontal, nor strictly vertical nor along a BZ edge. This is just a symbol indicating that the orientation of AWDv is not pinned to any particular direction. The actual orientation can be different than the one adopted in Figure \ref{Figu2}, although eq. (\ref{disT}) is used for presenting AWDv on the same star of the wave-vector. Dispersion and Hamiltonian are:
\numparts
\label{avDir}
\begin{equation}
\label{avDirD}
E_{1,2}=E_0\pm \sqrt{c_1q_1^2+c_2q_2^2+c_3q_1q_2},
\end{equation}
\begin{equation}
\label{avDirH}
\hat{H}=E_0\hat{\sigma}_0+\left(
\begin{array}[c]{cc}
	b_1q_1+b_2q_2 & z_1q_1+z_2q_2  \\
	z_1^*q_1+z_2^*q_2 & -b_1q_1-b_2q_2
\end{array}
\right),
\end{equation}
\endnumparts
where $c_1=b_1^2+|z_1|^2$, $c_2=b_2^2+|z_2|^2$, $c_3=2b_1b_2+z_1z_2^*+z_1^*z_2$. For groups $3^D$, $22^D$, $26^D$, $49^D$, $50^D$ and $73^D$, $b_1=b_2=0$. The parameters $c_1$, $c_2$ and $c_3$ do not satisfy eq. (\ref{uslov}).

\subsubsection{Anisotropic Weyl dispersion with fixed orientation (AWDf)}
\label{afDirDs}
AWDf consists of two anisotropic, spinful non-degenerate cones, whose orientation is unaltered by the symmetry preserving perturbations. It appears only when SOC is included and it is presented by a blue ellipse whose orientation is strictly vertical, strictly horizontal, along the BZ edge or perpendicular to it. Again, in actual orientation major and minor axes may interchange, and eq. (\ref{disT}) is used for presentation in Figure \ref{Figu2}. The dispersion can be reduced to:
\begin{equation}
\label{afDirD}
E_{1,2}=E_0\pm \sqrt{c_1q_1^2+c_2q_2^2}.
\end{equation}
Hamiltonians of interest can be derived from (\ref{avDirH}). For groups $8^D$, $9^D$, $10^D$, $11^D$, $12^D$, $13^D$, $31^D$, $32^D$, $33^D$ and $34^D$, $b_2=0$ and $z_1=0$ so that $c_1=b_1^2$ and $c_2=|z_2|^2$. For groups $19^D$, $20^D$, $22^D$, $23^D$, $24^D$, $26^D$, $53^D$, $55^D$, $57^D$ and $59^D$, $b_1=b_2=0$, $z_1=ib_3$ and $z_2=b_4$ so that $c_1=b_3^2$ and $c_2=b_4^2$. For groups $21^D$ and $25^D$, $b_2=0$, $z_1=0$ and $z_2=ib_3$ so that $c_1=b_1^2$ and $c_2=b_3^2$. For groups $67^D$, $68^D$, $69^D$ and $70^D$, $b_1=\sqrt{3}b_2$ and $z_2=-\sqrt{3}z_1$ so that $E_{1,2}=E_0\pm \sqrt{|z_1|^2(q_1-\sqrt{3}q_2)^2+b_2^2(q_2+\sqrt{3}q_1)^2}$. For groups $76^D$ and $77^D$, $b_1=b_2=0$ and $z_2=z_1^r/\sqrt{3}-i\sqrt{3}z_1^i$ so that $E_{1,2}=E_0\pm \sqrt{(z_1^i)^2(q_1-\sqrt{3}q_2)^2+(z_1^r)^2(q_1+q_2/\sqrt{3})^2}$. Last two dispersions are examples where eq. (\ref{uslov}) applies with $u=1/\sqrt{3}$.

\subsubsection{Fortune teller dispersion (FT)}
\label{FTs}
FT appears both in SOC and non-SOC cases, with the degeneracies being doubled in the latter \cite{Ja17, Mi20}. It is denoted by the quad arrow, yellow for non-SOC and red for SOC case. The dispersion is:
\numparts
\label{FT}
\begin{equation}
\label{FTd}
E_{1,2,3,4}=E_0\pm \left|\left|b_1q_2\right|\pm c\left|q_1\right|\right|.
\end{equation}
For layer double groups the Hamiltonian is:
\begin{equation}
\label{FTh}
\hat{H}=E_0\hat{I}_4+\left(
\begin{array}[c]{cccc}
	b_2q_1 & ib_1q_2 & 0 & zq_1 \\
	-ib_1q_2 &b_2q_1 &-zq_1 & 0 \\
	0 & -z^*q_1 &-b_2q_1 & ib_1q_2 \\
	z^*q_1 & 0 & -ib_1q_2 & -b_2q_1
\end{array}
\right),
\end{equation}
\endnumparts
so that $c=\sqrt{\left|z\right|^2+b_2^2}>0$ and $\hat{I}_4$ is the unit matrix. For layer single groups the total Hamiltonian is $\hat{H}_{tot}=\hat{\sigma}_0\otimes\hat{H}$, with $\hat{H}$ given by (\ref{FTh}) and $\left(\forall j\in\left\{1, 2, 3, 4\right\}\right) E_j^{\uparrow}=E_j^{\downarrow}=E_j$. For groups 43 and 45, $b_2=0$.

\subsubsection{Poppy flower dispersion (PF)}
\label{PF1s}
PF appears only in some non-centrosymmetric groups when SOC is included \cite{Mi20}. It is represented in Figure \ref{Figu2} by a red double-ellipse. Degeneracy at $\mathbf{q}=0$ is 4-fold (spin included) and it splits into spinful non-degenerate bands away from $q_1=0$ and $q_2=0$ lines. Dispersion and Hamiltonian are:
\numparts
\label{PF1}

\begin{equation}
\label{PFd}
E_{1,2,3,4}=E_0\pm\sqrt{c_1q_1^2+c_2q_2^2\pm c_3|q_1q_2|}.
\end{equation}
For groups $21^D$ and $25^D$:
\begin{equation}
\label{PF1H}
\hat{H}=E_0\hat{I}_4+\left(
\begin{array}[c]{cccc}
	b_1q_1 & b_2q_2 & z_1q_1 & z_2q_2 \\
	b_2q_2 & -b_1q_1 & z_2q_2 & -z_1q_1 \\
	z_1^*q_1 & z_2^*q_2 &-b_1q_1 & -b_2q_2 \\
	z_2^*q_2 & -z_1^*q_1 & -b_2q_2 & b_1q_1
\end{array}
\right),
\end{equation}
here $c_3=\sqrt{4b_1^2|z_2|^2+4b_2^2|z_1|^2-(z_1z_2^*-z_1^*z_2)^2-4b_1b_2(z_1z_2^*+z_1^*z_2)}$, $c_2=b_2^2+|z_2|^2$, $c_1=b_1^2+|z_1|^2$.
For groups $28^D$, $30^D$, $32^D$ and $34^D$:
\begin{equation}
\label{PF2H}
\fl \hat{H}=E_0\hat{I}_4+\left(
\begin{array}[c]{cccc}
	b_1q_1+b_2q_2 & 0 & z_1q_1+z_2q_2 & 0 \\
	0 & b_1q_1-b_2q_2 & 0 & z_1q_1-z_2q_2 \\
	z_1^*q_1+z_2^*q_2 & 0 &-b_1q_1-b_2q_2 & 0 \\
	0 & z_1^*q_1-z_2^*q_2 & 0 & -b_1q_1+b_2q_2
\end{array}
\right),
\end{equation}
here $c_1=b_1^2+|z_1|^2$, $c_2=b_2^2+|z_2|^2$, $c_3=|2b_1b_2+z_1z_2^*+z_1^*z_2|$.
For groups $54^D$, $56^D$, $58^D$ and $60^D$:
\begin{eqnarray}
\label{PF3H}
\fl &&\hat{H}=E_0\hat{I}_4 \\
\fl &+&\left(
\begin{array}[c]{cccc}
	0 & (1-i)b_1(q_1-iq_2) & z^*(q_1+iq_2) & 0 \\
	(1+i)b_1(q_1+iq_2) & 0 & 0 & z^*(q_2+iq_1) \\
	z(q_1-iq_2) & 0 & 0 & (1-i)b_1(q_2-iq_1) \\
	0 & z(q_2-iq_1) & (1+i)b_1(q_2+iq_1) & 0
\end{array}
\right), \nonumber
\end{eqnarray}
here $c_1=c_2=2b_1^2+|z|^2$, $c_3=4\sqrt{2}|b_1z|$ (called isotropic PF in \cite{Mi20}).
\endnumparts

\subsubsection{Anisotropic Dirac dispersion with variable orientation (ADDv)}
\label{DDvs}
ADDv is analogous to AWDv with degeneracy of each band doubled. It appears only in SOC case. ADDv is represented by a red ellipse with the same convention as for AWDv described in Section \ref{avDirs}. Dispersion and Hamiltonian are:
\numparts
\label{DDv}
\begin{equation}
\label{DDvD}
E_{1,2,3,4}=E_0\pm\sqrt{c_1q_1^2+c_2q_2^2+c_3q_1q_2},
\end{equation}
\begin{equation}
\label{DDvH}
\fl \hat{H}=E_0\hat{I}_4+\left(
\begin{array}[c]{cccc}
	0 & b_1q_1+b_2q_2 & z_1q_1+z_2q_2 & 0 \\
	b_1q_1+b_2q_2 & 0 & 0 & z_1q_1+z_2q_2 \\
	z_1^*q_1+z_2^*q_2 & 0 &0 & -b_1q_1-b_2q_2 \\
	0 & z_1^*q_1+z_2^*q_2 & -b_1q_1-b_2q_2 & 0
\end{array}
\right),
\end{equation}
\endnumparts
here $c_1=b_1^2+|z_1|^2$, $c_2=b_2^2+|z_2|^2$, $c_3=2b_1b_2+z_1z_2^*+z_1^*z_2$.

\subsubsection{Anisotropic Dirac dispersion with fixed orientation (ADDf)}
\label{DDfs}
ADDf is analogous to AWDf with degeneracy of each band doubled. It appears only in SOC case and it is represented by a red ellipse. The convention follows the one for AWDf in Section \ref{afDirDs}. Dispersion and Hamiltonian are:
\numparts
\label{DDf}
\begin{equation}
\label{DDfD}
E_{1,2,3,4}=E_0\pm\sqrt{c_1q_1^2+c_2q_2^2},
\end{equation}
For groups $15^D$, $16^D$ and $17^D$:
\begin{equation}
\label{DDfH}
\hat{H}=E_0\hat{I}_4+\left(
\begin{array}[c]{cccc}
	b_2q_2 & ib_1q_1 & zq_2 & 0 \\
	-ib_1q_1 & -b_2q_2  & 0 & -zq_2 \\
	z^*q_2 & 0 &-b_2q_2 & ib_1q_1 \\
	0 & -z^*q_2 & -ib_1q_1 & b_2q_2
\end{array}
\right),
\end{equation}
here $c_1=b_1^2$, $c_2=b_2^2+|z|^2$. For groups $38^D$, $39^D$(points $\left(\pm \pi/a_1,0\right)$ and $\left(0, \pm \pi/a_2\right)$), $41^D$, $42^D$, $43^D$, $45^D$, $46^D$(points $\left(\pm \pi/a_1,0\right)$ and $\left(0, \pm \pi/a_2\right)$), $62^D$(points $\left(\pm \pi/a,0\right)$ and $\left(0, \pm \pi/a\right)$) and $64^D$(points $\left(\pm \pi/a,0\right)$ and $\left(0, \pm \pi/a\right)$), $b_2=0$. For groups $39^D$(BZ-corners) and $46^D$(BZ-corners), $z=0$.
\endnumparts

\subsubsection{Isotropic Dirac dispersion (IDD)}
\label{IsoDDs}
IDD is analogous to IWD with degeneracy of each band doubled. It appears in cases with and without SOC. An example of IDD is dispersion near $K$-points in graphene. IDD is represented by a red circle (when no other dispersions near this point are possible) or by a green ring (when SB is an alternative). Dispersion and Hamiltonian are:
\numparts
\label{IsoDD}
\begin{equation}
\label{IsoDDD}
E_{1,2,3,4}=E_0\pm|z||\mathbf{q}|
\end{equation}
For group $52^D$:
\begin{equation}
\label{IsoDDH}
\fl \hat{H}=E_0\hat{I}_4+\left(
\begin{array}[c]{cccc}
	0 & 0 & z(q_1-iq_2) & 0 \\
	0 & 0 & 0 & z(q_1-iq_2) \\
	z^*(q_1+iq_2) & 0 & 0 & 0 \\
	0 & z^*(q_1+iq_2) & 0 & 0
\end{array}
\right),
\end{equation}
\endnumparts
for groups $62^D$ and $64^D$, $z=ib_1$. For layer single groups $\hat{H}_{tot}=\hat{\sigma}_0\otimes\hat{H}$, with $\hat{H}$ given by (\ref{DirH}) and $E_1^{\uparrow}=E_1^{\downarrow}=E_1=E_2$, $E_2^{\uparrow}=E_2^{\downarrow}=E_3=E_4$. For layer single groups $z=b_1$ except for 66, 73 and 75.

\subsection{Quadratic dispersions}

\subsubsection{Corner quadratic dispersion (CQ)}
\label{kvads}
CQ appears at BZ corners of rectangular groups where two DNL meet. It appears both in SOC and non-SOC cases and it is denoted by a red rectangle. The 4-fold spinfull degenerate band at $\mathbf{q}=0$, splits into two double degenerate bands for $\mathbf{q}\neq 0$ and away from DNLs. The splitting causing term is of the second order (quadratic) in $\mathbf{q}$. The dispersion is:
\numparts
\label{kvad}
\begin{equation}
\label{kvadD}
E_{1,2,3,4}=E_0+a_1q_1^2+a_2q_2^2\pm c\left|q_1q_2\right|.
\end{equation}
For layer single groups $E_1^{\uparrow}=E_1^{\downarrow}=E_1=E_2$, $E_2^{\uparrow}=E_2^{\downarrow}=E_3=E_4$ and the Hamiltonian is:
\begin{equation}
\label{kvadH}
\hat{H}=\hat{\sigma}_0\otimes\left[(E_0+a_1q_1^2+a_2q_2^2)\hat{\sigma}_0+q_1q_2\left(b_1\hat{\sigma}_1+b_2\hat{\sigma}_2\right)\right],
\end{equation}
so that $c=\sqrt{b_1^2+b_2^2}$. For groups 32, 34, 39, 42, 46, $b_2=0$. For layer double groups:
\begin{equation}
\label{sockvH}
\hat{H}=(E_0+a_1q_1^2+a_2q_2^2)\hat{I}_4+q_1q_2\left(
\begin{array}[c]{cccc}
	0 & ib_1 & 0 & z \\
	-ib_1 & 0 & -z & 0 \\
	0 & -z^* & 0 & -ib_1 \\
	z^* & 0 & ib_1 & 0
\end{array}
\right),
\end{equation}
\endnumparts
so that $c=\sqrt{b_1^2+|z|^2}$.

\subsubsection{Isotropic corner quadratic dispersion (ICQ)}
\label{iskvads}
In the BZ corners of square groups where two DNL meet, the axis of order four imposes $a_1=a_2$ in CQ and gives ICQ. The term \emph{isotropic} denotes that disperions along $q_1=0$ and $q_2=0$ are identical. ICQ appears in both SOC and non-SOC cases and it is marked by a red square. The dispersion is:
\numparts
\label{iskvad}
\begin{equation}
\label{iskvadD}
E_{1,2,3,4}=E_0+aq^2\pm c\left|q_1q_2\right|.
\end{equation}
For layer single groups $E_1^{\uparrow}=E_1^{\downarrow}=E_1=E_2$, $E_2^{\uparrow}=E_2^{\downarrow}=E_3=E_4$ and the Hamiltonian is:
\begin{equation}
\label{iskvadH}
\hat{H}=\hat{\sigma}_0\otimes\left[(E_0+aq^2)\hat{\sigma}_0+q_1q_2\left(b_1\hat{\sigma}_1+b_2\hat{\sigma}_2\right)\right],
\end{equation}
\endnumparts
so that $c=\sqrt{b_1^2+b_2^2}$. For correps that remain irreducible when TRS becomes broken, $b_2=0$. For layer double group $63^D$, $a_1=a_2=a$ and $b_1=0$ in (\ref{sockvH}) so that $c=|z|$. 

\subsubsection{Quadratic dispersion (QD)}
\label{cukvs}
QD appears near isolated BZ points only in non-SOC case \cite{HONP}. It is denoted by the green star in Figure \ref{Figu1}, since SB can appear instead. The splitting in all directions away from the point is quadratic, as indicated by dispersion and Hamiltonian:
\numparts
\label{cukv}
\begin{equation}
\label{cukvD}
\fl E_{1,2}^{\uparrow}=E_{1,2}^{\downarrow}=E_0+aq^2\pm\sqrt{[b_3(q_1^2-q_2^2)+b_1q_1q_2]^2+[b_4(q_1^2-q_2^2)+b_2q_1q_2]^2},
\end{equation}
\begin{equation}
\label{cukvH}
\fl \hat{H}=\hat{\sigma}_0\otimes\left\{(E_0+aq^2)\hat{\sigma}_0+[b_3(q_1^2-q_2^2)+b_1q_1q_2]\hat{\sigma}_1+[b_4(q_1^2-q_2^2)+b_2q_1q_2]\hat{\sigma}_2\right\}.
\end{equation}
\endnumparts
Except for groups 49, 50, 51 and 52, in (\ref{cukvD}) and (\ref{cukvH}) $b_1=b_4=0$. If $b_1b_4=b_2b_3$, two lines of accidental degeneracy intersect at $\mathbf{q}=0$ symmetrically under rotation by $\pi/2$, so that $\mathbf{q}=0$ is not an isolated point any more.

\subsubsection{Isotropic quadratic dispersion (IQD)}
\label{kvs}
IQD is QD which is fully isotropic. It appears only in non-SOC case \cite{HONP} and it is indicated with a green snowflake, since the appearance of SB is other possibility. Dispersion and Hamiltonian are:
\numparts
\label{kv}
\begin{equation}
\label{kvD}
E_{1,2}^{\uparrow}=E_{1,2}^{\downarrow}=E_0+\left(a\pm \sqrt{b_1^2+b_2^2}\right)q^2,
\end{equation}
\begin{equation}
\label{kvH}
\fl \hat{H}=\hat{\sigma}_0\otimes\left\{(E_0+aq^2)\hat{\sigma}_0+[2b_2q_1q_2+b_1(q_1^2-q_2^2)]\hat{\sigma}_1+[b_2(q_1^2-q_2^2)-2b_1q_1q_2]\hat{\sigma}_2\right\},
\end{equation}
\endnumparts
with $b_2=0$ except for groups 65, 66, 73, 74 and 75. Unlike QD, IQD can not be turned into a line of degeneracies by suitable choice of non-zero parameters.

\subsubsection{Simple band (SB)}
\label{jdkvs}
SB is a band (spinful non-degenerate, with SOC and double spinful degenerate in non-SOC case) which disperses quadratically near $\mathbf{q}=0$ without any splitting (\emph{i.e.} it is neither UNP nor UNL). It is introduced here only when it appears as an alternative to other dispersions. These cases are denoted by green symbols in Figures \ref{Figu1} and \ref{Figu2}. The dispersion and Hamiltonian are:
\numparts
\label{jdkv}
\begin{equation}
\label{jdkvD}
E_1=E_0+aq^2,
\end{equation}
\begin{equation}
\label{jdkvH}
\hat{H}=E_0+aq^2.
\end{equation}
\endnumparts
For layer single groups $\hat{H}_{tot}=(E_0+aq^2)\hat{\sigma}_0$ and $E_1^{\uparrow}=E_1^{\downarrow}=E_1$. 

\subsection{Cubic dispersions}

\subsubsection{Cubic disperson (CD)}
\label{cps}
CD appears only when SOC is included \cite{HONP}. It consists of spinful double degenerate band at $\mathbf{q}=0$ which splits cubically away from this point. It is denoted with a green square, since IWD is other dispersion possible at groups/BZ points form Figure \ref{Figu2}. Dispersion and Hamiltonian are:
\numparts
\label{cp}
\begin{eqnarray}
\label{cpD}
\fl E_{1,2}&&=E_0+a|\mathbf{q}|^2 \nonumber \\
\fl &&\pm \sqrt{\left|z_1q_1(q_1^2-3q_2^2)+z_2q_2(q_2^2-3q_1^2)\right|^2+[b_1q_1(q_1^2-3q_2^2)+b_2q_2(q_2^2-3q_1^2)]^2},
\end{eqnarray}
\begin{eqnarray}
\label{cpH}
\fl \hat{H}&=&(E_0+a|\mathbf{q}|^2)\hat{\sigma}_0 \nonumber \\
\fl &+&\left(
\begin{array}[c]{cc}
	b_1q_1(q_1^2-3q_2^2)+b_2q_2(q_2^2-3q_1^2) & z_1q_1(q_1^2-3q_2^2)+z_2q_2(q_2^2-3q_1^2)  \\
	z_1^*q_1(q_1^2-3q_2^2)+z_2^*q_2(q_2^2-3q_1^2) & -b_1q_1(q_1^2-3q_2^2)-b_2q_2(q_2^2-3q_1^2)
\end{array}
\right).
\end{eqnarray}
\endnumparts
If vectors $\left\{(z_1^r,z_2^r)^T, (z_1^i,z_2^i)^T\right\}$ are linearly dependent and vectors $\left\{(z_1^r,z_2^r)^T, (b_1,b_2)^T\right\}$ are also linearly dependent, then three accidental nodal lines intersect symmetrically under rotation by $\pi/3$ at $\mathbf{q}=0$, so that $\mathbf{q}=0$ is not an isolated point any more. For groups $67^D$, $68^D$, $69^D$ and $70^D$, $b_1=0$ and $z_2=0$. For group $73^D$, $b_1=b_2=0$. For groups $76^D$ and $77^D$, $b_1=b_2=0$, $z_1=ib_3$ and $z_2=b_4$.

\subsubsection{Triangular cubic dispersion (TC)}
\label{hexas}
TC appears only in SOC case, when three WNL intersect at a single point. It is denoted by a blue triangle in Figure \ref{Figu2}. The splitting is caused by cubic terms in the Taylor expansion of Hamiltonian. Dispersion and Hamiltonian are:
\numparts
\label{hexa}
\begin{equation}
\label{hexaD}
E_{1,2}=E_0+aq^2\pm \left|b\right|\left|q_2(q_2^2-3q_1^2)\right|,
\end{equation}
\begin{equation}
\label{hexaH}
\hat{H}=(E_0+aq^2)\hat{\sigma}_0+b\:\!q_2(q_2^2-3q_1^2)\hat{\sigma}_3.
\end{equation}
\endnumparts
TC appears in Hexagonal groups $78^D$ and $79^D$.

\section{Discussion and conclusions}
Comparison of Figs. \ref{Figu1} and \ref{Figu2}, allows conclusions on how \emph{e.g.} inclusion of SOC affects dispersions from Figure \ref{Figu1}. Dispersions in groups 40, 44 and 63 are unaltered, up to numerical values of parameters, by the inclusion of SOC. In groups 43 and 45, DNLs perpendicular to $\mathbf{k}_1$ are also preserved when SOC is included. On the other hand, SOC induces DNL to split into two WNLs in groups 9, 12, 20, 21, 24, 25, 28, 29, 30, 32 (DNL perpendicular to $\mathbf{k}_1$), 33 (DNL perpendicular to $\mathbf{k}_1$), 34 (DNL perpendicular to $\mathbf{k}_1$), 54, 56, 58 and 60. In groups 21 and 25, SOC induces transition from CQ to PF, while in square groups 54, 56, 58 and 60, SOC turns ICQ to isotropic PF. Other SOC-induced transitions are CQ to ADDf (groups 39 and 46) and ICQ to IDD in groups 62, 64. SOC induced splittings at HSP in non-magnetic layers are also discussed in \cite{Mi22}.

Dispersions presented in Section \ref{Res} are exact, in the sense that if we knew how to solve the one-electron Schroedinger equation for a single crystal exactly, we would get the very same formulas (with the numerical values of parameters as an additional result). The dispersions are unaltered when electron correlations are included, as long as they are weak enough as to preserve the quasiparticle picture. For these reasons, each dispersion presented here can be a research topic of its own. The synthesis of a new 2D material with the right placement of the Fermi level, would likely trigger new research directions.

Besides band contacts treated here, there are accidental band contacts which appear along HSLs or away from them in the BZ. Some of these contacts are guaranteed to exist by the topological laws. Such guaranteed contacts can be moved but not removed by the symmetry preserving perturbations and are tabulated for (3D) hexagonal \cite{AccBCh}, tetragonal \cite{AccBCt} and orthorhombic \cite{AccBCo} space groups. These results could be used for layer groups, by subduction from corresponding space groups to 2D BZ. On the other hand, truly accidental band contacts can be destroyed by a perturbation that keeps the symmetry, which in real materials can be caused by the change of temperature, pressure, strain \emph{etc}. Four-fold spinless degenerate points that belong to HSL, treated in \cite{OsKin, OsKinE, DirPho}, are examples of truly accidental band contacts.

Experimental realization of a 2D material with prescribed layer symmetry is far from being trivial, so one could search across database of 3D materials which have already been synthesized and exfoliate easy into layers \cite{Exp}. It contains materials of almost all 3D symmetries obtained by periodic repetition of layers perpendicularly. Dispersions presented in Section \ref{Res} are of significance if in a 2D material the Fermi level crosses bands near $E_0$. Although this cannot be guaranteed solely by group-theoretical arguments, there are some symmetry based results which are helpful. Refs. \cite{ElFil1, ElFil2, ElFil3} give electron fillings necessary for a crystal belonging to certain space or layer group to be insulating. When not fulfilled, the crystal is metallic (or insulating if the electron-electron interaction is strong). For example, if in a 2D material belonging to layer group 33, the sum of atomic numbers of all nuclei in the primitive cell is not divisible by eight, the material is a metal \cite{ElFil1, ElFil2, ElFil3}. It would be possible then to dope the material with electrons or holes by \emph{e.g.} gating, especially if the material belongs to layer group which is also a symmetry of a surface (a wallpaper group).

%The code used was:
%\small\begin{verbatim}
%\begin{equation}
%\label{cases}
%X=\cases{1&for $x \ge 0$\\
%-1&for $x<0$\\}
%\end{equation}
%\end{verbatim}
%\normalsize
\ack{Authors acknowledge funding by the Ministry of Science, Technological Development and Innovation of the Republic of Serbia provided by the Institute of Physics Belgrade (V.D.) and Faculty of Physics (N.L.).}

%%%%\bibliography{Ref2212}
\bibliographystyle{unsrt}

\end{document}